\documentclass{article}
\usepackage[utf8]{inputenc}
\usepackage{graphicx}
\usepackage{url}

\title{Muon Colliders:\\
Opening New Horizons for Particle Physics} 
\author{
K. Long, D. Lucchesi, M. Palmer, N. Pastrone, D. Schulte, V.Shiltsev}
\date{\today{}}

\begin{document}
\maketitle
\begin{abstract}
  \noindent
  Particle colliders have arguably been the most important
  instruments for particle physics over the past 50 years.
  As they became more powerful, they were used to push the frontier of our
  knowledge into previously uncharted territory. The LHC, the highest
  energy collider to date, at which the Higgs boson was discovered, is a
  prime example.
  To continue along the road into the {\it Terra Promissa} beyond the
  Standard Model requires colliders with energy reach even greater
  than that of the LHC.  
  Beams of muons offer enormous
  potential for the exploration of the energy frontier. Since the muon
  is a fundamental particle, its full energy is available in collisions in contrast to protons which are composed of quarks and gluons.  
However, muon beams decay rapidly, which presents a special challenge for a collider. Recent research
indicates that the technologies required to overcome this challenge are
within our grasp and may offer a cost-effective and energy-efficient option to continue our explorations.
  A new international collaboration is forming to bring
  together the diverse expertise and complementary capabilities from
  around the world to realise the muon collider as the next-generation energy-frontier discovery machine.

\end{abstract}

\noindent
Evaluation of future high energy particle physics facilities has
traditionally been based on three criteria: scientific potential; technical construction and financial
requirements; and flexibility for further upgrades and
developments~\cite{martinirvin1984}. 
The most recent update of the European Strategy for Particle
Physics has added an important new 
requirement---next generation facilities should meet very high 
ecological and environmental standards and, in particular, they 
must be energy efficient ~\cite{eppsu2020}.

Colliding two particle beams, each with energy $E$, allows the
exploration of center-of-mass energy scales up to $\sqrt{s}=2E$.
Since they were developed in the 1960s, colliders have been built in a variety of types, shapes, and sizes.
Today, colliders represent the largest and most expensive facilities for fundamental science research~\cite{Shiltsev2020rmp}.
To fulfill our aspirations of an order-of-magnitude increase in the
center-of-mass energy in particle collisions requires a paradigm shift from the traditional and well-established technologies of
proton-proton and electron-positron colliders.

\begin{figure}
  \begin{center}
    \includegraphics[width=0.95\textwidth]{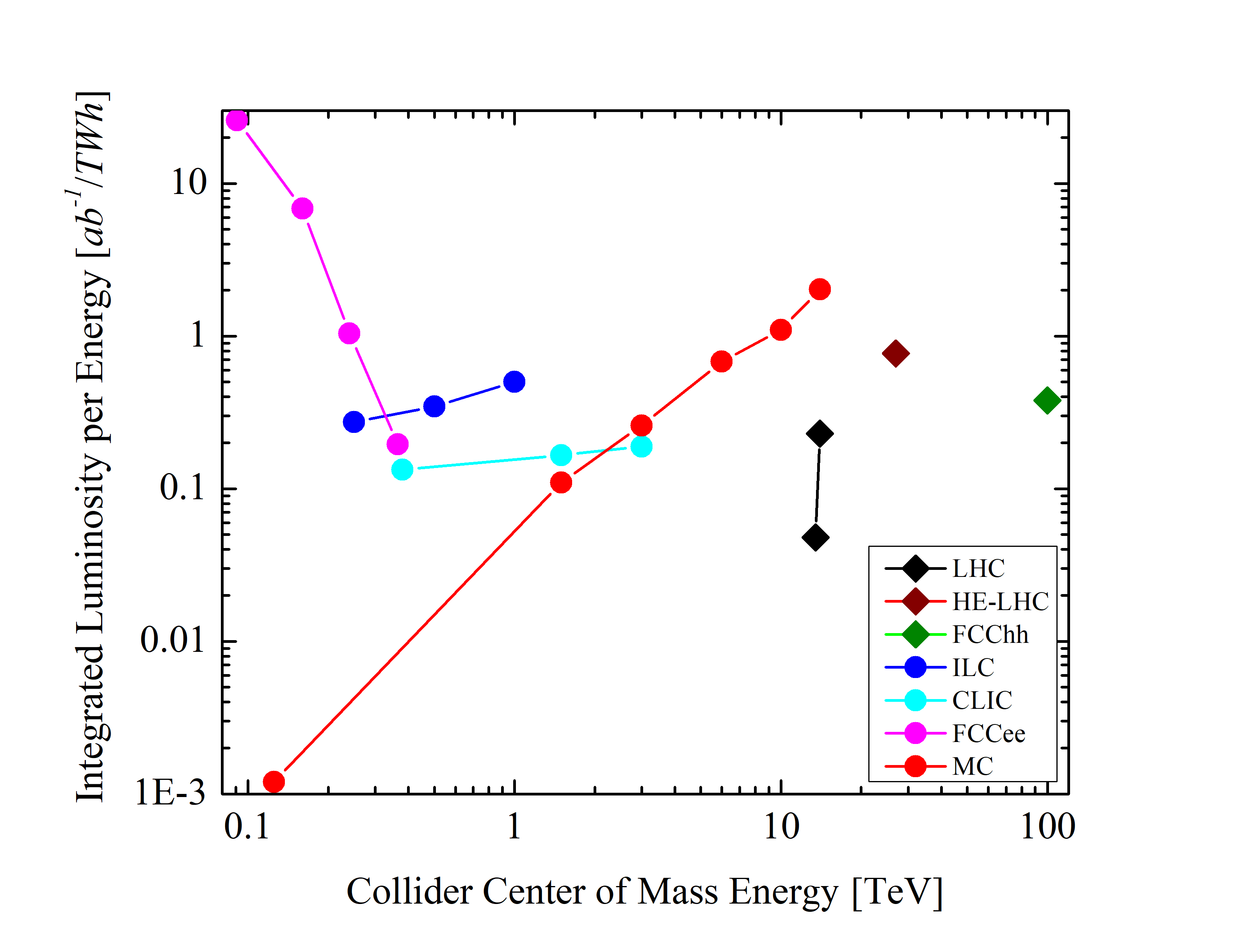}
  \end{center}
  \caption{
    Annual integrated luminosity, $\int L dt$, per Terawatt-hour of electric power
    consumption, $\int P dt$, plotted as a function of the center-of-mass energy for a variety of particle colliders, as taken from reference~\cite{Shiltsev2020rmp}.
    This ratio is shown as a function of center-of-mass energy for: the Muon Collider (MC) (solid red circles); the LHC, both at present and after
    its luminosity upgrade (solid black diamonds); the Future Circular  electron-positron Collider (FCC-ee, solid magenta circles) assuming experiments at two collision points; the International Linear Collider (ILC, solid blue circles), the Compact Linear Collider (CLIC, solid cyan circles), the High Energy LHC (HE-LHC, solid magenta diamonds); and the Future Circular proton-proton Collider (FCC-hh, solid green diamonds). 
    The effective energy reach of hadron colliders (LHC, HE-LHC and
    FCC-hh) is approximately a factor of 7 lower than that of a lepton
    collider operating at the same energy per beam. 
  }  
  \label{ABperTWH}
\end{figure}

The concept of colliding beams of positively and negatively charged
muons originated in the late 1960s~\cite{tikhonin1968}. Two fundamental issues drive the design of a muon collider---the short lifetime of the muon and the challenge of producing bright muon beams.  
Detailed design studies and technological
research~\cite{MC1999,geer2009,palmer2014muonrast,boscolo2018futuremuon, mapc}
have demonstrated the feasibility of using muon beams in 
a multi-TeV collider that addresses each of the criteria set out
in the 2020 update of the European Strategy for Particle Physics.
Muons can be accelerated and collided in rings without suffering from the large synchrotron radiation losses that limit the performance of electron-positron colliders.
This allows a muon collider to use the traditional and well-established accelerator technologies of superconducting high-field magnets and RF cavities.
Furthermore, unlike protons, where the collision energy is shared among the constituent quarks and gluons, muons are point-like particles that deliver their entire energy $\sqrt{s}$ to the collision - thus probing significantly higher energy scales  ($\sim 7\times$ higher, depending on the process) than protons colliding with the same beam energy.
Finally, the electric power efficiency of muon colliders (defined
as the collider's annual integrated luminosity, $\int L dt$, divided by the
facility's annual energy use) increases with energy.
Thus, above approximately $\sqrt{s}=2$\,TeV, a muon collider is expected to be 
the most energy-efficient choice for the exploration of the energy
frontier as shown in Figure~\ref{ABperTWH}.
This would be the first time unstable particles are ussed in a collider.

\section{Physics at a Multi-TeV Muon Collider}

\begin{figure}
  \begin{center}
    \includegraphics[width=0.75\textwidth]{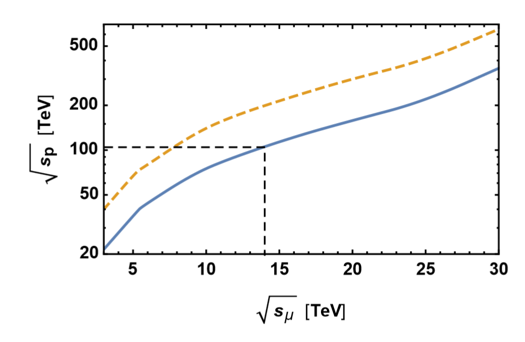}
  \end{center}
  \caption{
    Energy reach of muon-muon collisions: the energy at which the
    proton collider cross-section equals that of a muon collider
    (taken from reference~\cite{ESInput}).
    The plot compares the pair-production cross-sections for heavy
    particles with mass $M$ at approximately half the muon
    collider energy $\sqrt{s_\mu}/2$. 
    The dashed yellow line assumes comparable processes for muon and
    proton production, while the continuous blue line accounts for the
    possible QCD enhancement of the production rates at a proton-proton
    collider.
  } 
  \label{MuonProton}
\end{figure}

A muon collider operating at an energy $>$10\,TeV
would enable direct searches for new particles over a wide range of unexplored masses \cite{Barger}.  Figure~\ref{MuonProton} compares the center-of-mass energy $\sqrt{s_{\mu}}$ at a muon collider to that at a proton-proton machine, $\sqrt{s_{p}}$, required to directly produce a pair of new heavy particles of mass $M\sim\sqrt{s_{\mu}}/2$ with roughly the same probability in each collision. For these direct searches, a 14\,TeV muon 
collider with sufficient luminosity would provide similar discovery reach as a 100\,TeV proton-proton collider~\cite{ESInput}.
Many other physics models beyond the Standard Model could also be probed at such a high energy machine. For example, a multi-TeV muon collider would be a factory for production of pairs of the so-called super-symmetric particles. 
At energies greater than a few TeV, the dominant production process in muon collisions is electroweak vector boson fusion/scattering.  This process could be exploited for the detailed study of all Standard Model processes and would provide a precision probe for the appearance of new physics processes \cite{VBF}.

\begin{figure}
  \begin{center}
    \includegraphics[width=0.95\textwidth]{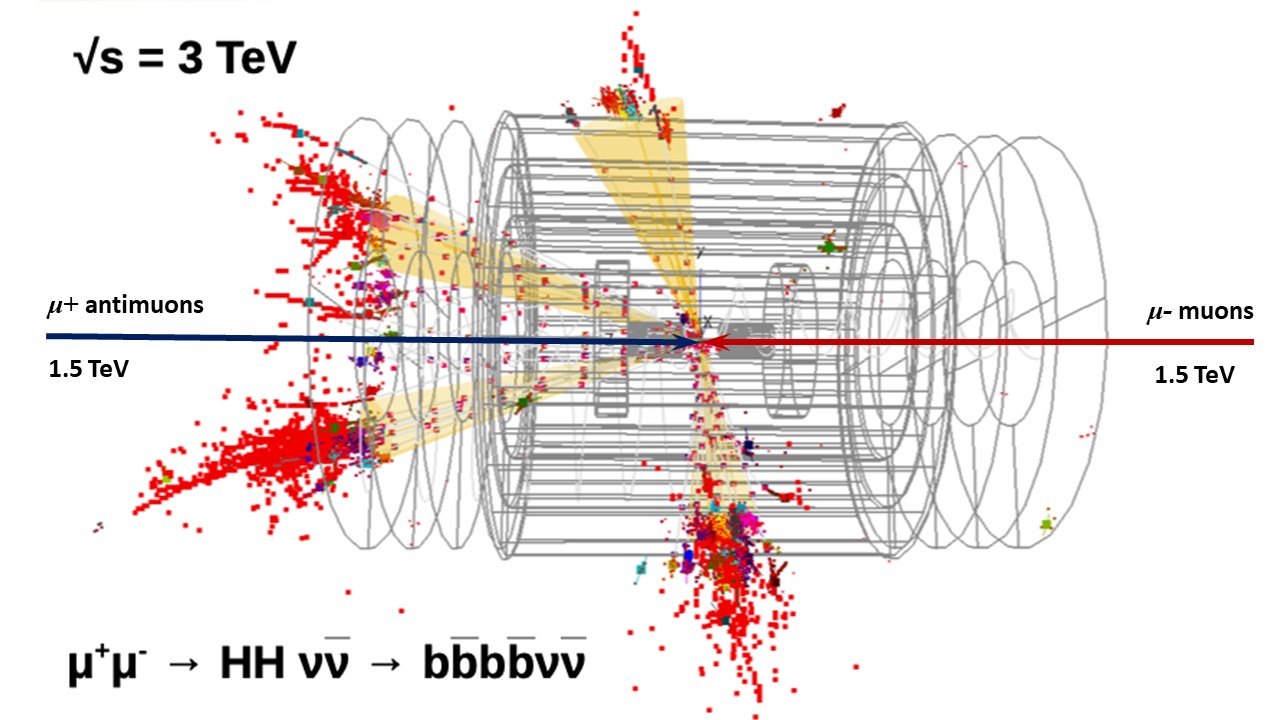}
  \end{center}
  \caption{
    A typical double Higgs boson event simulation produced
    at a center-of-mass energy of 3 TeV. Each Higgs boson is identified by its decay to a b and anti-b quark jet pair.
    Grey lines show the path in the tracking detectors, where charged particles produced by each b-jet are enclosed in a yellow cone. The red dots represent the jet shower energy deposition inside the calorimeter in the outer region of the detector.
  }  
  \label{HH3TeV}
\end{figure}

Muon collisions at a center-of-mass energies above 10\,TeV will enable fundamental investigations of Nature through the detailed study of the properties of the Higgs boson. With a luminosity of $\sim10^{35}$\,cm$^{-2}$ s$^{-1}$, it would be possible to produce enough double and triple-Higgs events to make a direct measurement of the trilinear and quadrilinear self-couplings of the Higgs boson, thus providing a unique and precise determination of the shape of the Higgs potential. Figure \ref{HH3TeV} shows how a typical double-Higgs event would appear in the detector when both Higgs bosons decay to b and anti-b quark jet pairs.

The detector must be capable of operating in the presence of the beam-induced background produced tens of meters upstream of the interaction point along the beam line by the interactions between the decay products of the muon beams and the machine elements. 
The particle types (mainly photons, electrons and neutrons), flux, angular, and energy distributions of the background all depend strongly on the exact details of the machine lattice. This requires the design of the machine-detector interface to be optimized along with the collider design at a given energy. Two tungsten shielding cones (nozzles), emanating from the collision point and inserted inside the tracker detector volume, mitigate the effects of the high levels of beam-induced background close to the beam pipe. The experiment, in  particular the  tracking system shown in Figure \ref{HH3TeV}, requires detectors with performance that exceeds the present state of the art, e.g., being capable of few tens of ps timing resolution to reject out-of-time beam-induced background~\cite{bartosik2019mcneutrinoradiation}. Detector designs must be developed further to enable simultaneous measurements of the position, time and energy of the particles originating from the collision point, as well as to exploit new artificial intelligence on-detector data handling and reconstruction tools.

\section{Main Collider Parameters, Systems and Synergies}
\label{muoncolliders}

\begin{figure}
  \begin{center}
    \includegraphics[width=1.0\textwidth]{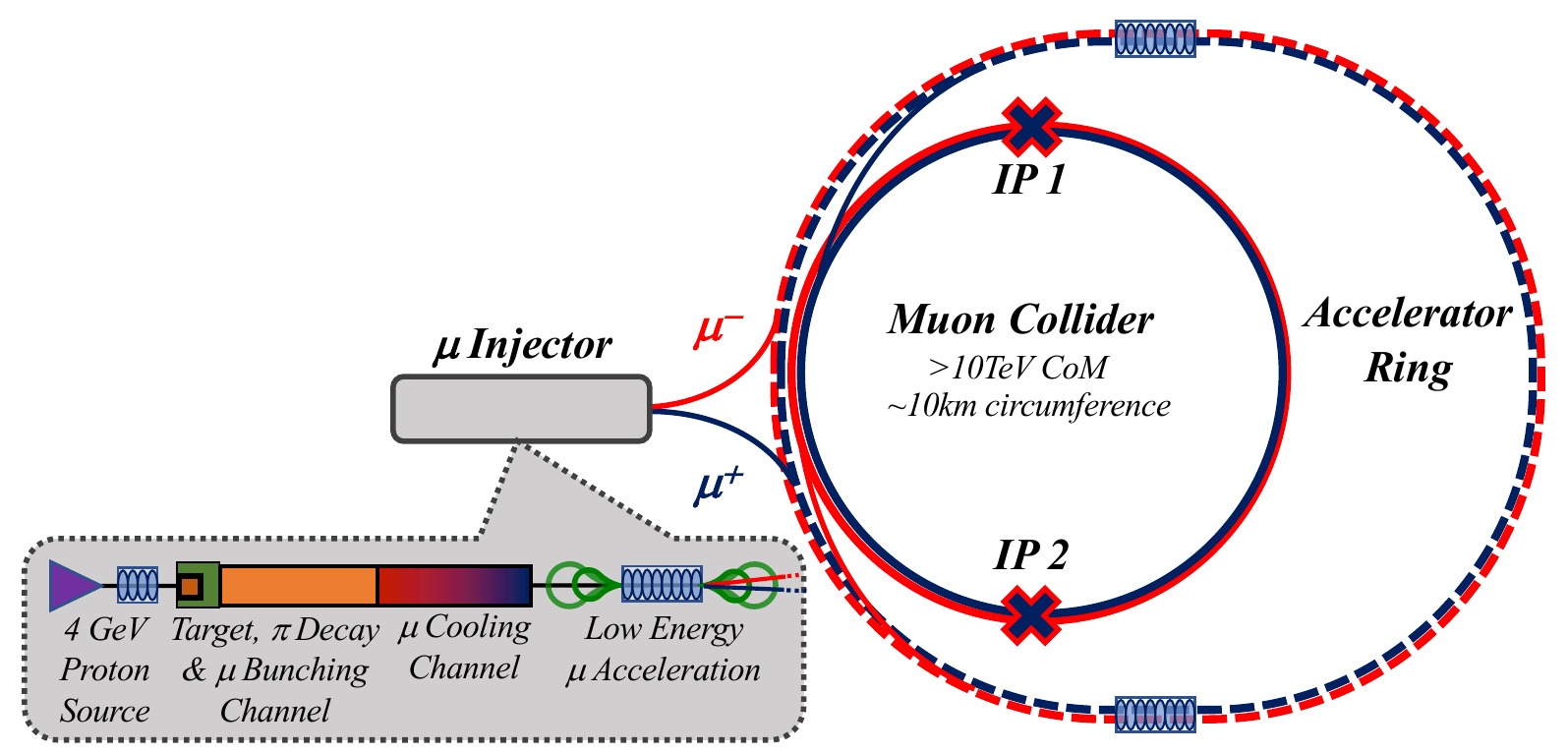}
  \end{center}
  \caption{
    Schematic layout of a 10 TeV-class muon collider complex.  The muon injector systems include the proton driver, a high power target system with capture solenoid for the pions generated by the proton interactions with the target, a pion decay channel where muons are collected and subsequently bunched together, a muon ionization cooling channel that provides cooling for both positive and negative muon beams by more than 5 orders of magnitude, and a low energy muon accelerator stage that would deliver beams with energies up to 100\,GeV.  From the injector, each species of muon beam is transferred into a high energy accelerator complex that can take the beams to the multi-TeV energies required.  Finally, the beams will be transferred to a smaller collider ring whose circumference is optimized for luminosity performance.  A 10 TeV-class collider ring is anticipated to support at least 2 detector interaction regions for the physics program. 
  }   
  \label{MCatCERN}
\end{figure}

As noted above, the short lifetime of the muon and the provision of a muon source with sufficient brightness represent particular challenges for the realisation of a muon collider.  
Rapid acceleration to high energies increases the lifetime of the
muons in proportion to their energy---a lifetime of 21\,ms is achieved
at 1\,TeV.
A possible layout for a 6--14\,TeV center-of-mass energy muon collider
with a luminosity $O$($10^{35}$~cm$^{-2}$s$^{-1}$) is shown in
Figure~\ref{MCatCERN}.
A high-power proton driver produces a dense beam that is further
compressed in an accumulator and a compressor ring.
The proton beam is directed onto a target where positive and negative
pions are produced that subsequently decay into positive and negative
muons.
The muons are captured in two beams, depending on their charge, that
are  subsequently cooled using the novel ionization-cooling technique
and then accelerated in several stages before being injected into the
collider ring.
The cooling system is essential to reduce the large initial phase
space by a factor of more than $10^5$ to that needed for a
practical collider. 
Without such cooling, the luminosity would be too small by several orders of magnitude.

\begin{figure}
  \begin{center}
    \includegraphics[width=0.95\textwidth]{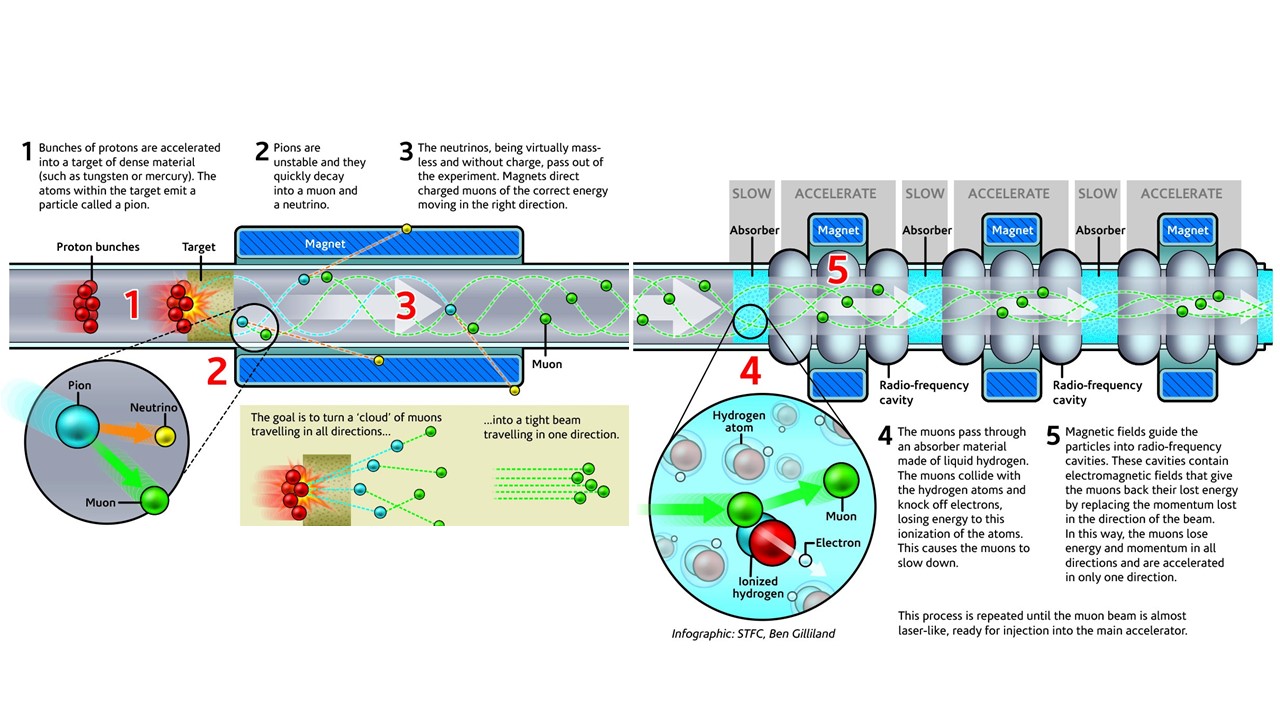}
  \end{center}
  \caption{
    Ionization cooling-channel scheme.
    200\,MeV muons are born when 4\,GeV protons hit a solid target.
    The muons are collected and sent into a cooling channel where they
    lose energy in lithium hydride (LiH) absorbers.
    The lost energy is restored when the muons are accelerated in the
    longitudinal direction in RF cavities.
    The superconducting solenoids magnets confine the beam
    within the channel and radially focus the beam at the absorbers.
  }  
  \label{MuCooling}
\end{figure}

The proposed ionization-cooling technique is very fast and uniquely
applicable to muons because of their minimal interaction with matter.
Ionization cooling involves passing the muon beam through some
material in which the muons lose energy via the ionization
energy-loss mechanism.  
Both transverse and longitudinal momentum are reduced in this process.
Longitudinal momentum only is then restored by subsequent
acceleration in RF cavities.
The combination of energy loss and re-acceleration causes a net
reduction in the phase space occupied by the beam, hence cooling the beam. 
The process is repeated many times to achieve a large cooling
factor--see Figure~\ref{MuCooling}.

Theoretical studies and numerical simulations\,\cite{palmer2014muonrast} have shown that, with realistic parameters for the cooling hardware, ionization cooling is expected to reduce the phase-space volume occupied by the
initial muon beam sufficiently to provide acceptable luminosity in a collider.
A complete cooling channel would consist of a series of tens of 
cooling stages, each reducing the 6D phase-space volume by a roughly a factor of
2 (see Figure~\ref{MuCooling}).
The ionization cooling method, though relatively straightforward in
principle, presents challenges in its practical implementation.
These challenges include the suppression of RF breakdown
normal-conducting RF cavities immersed in strong magnetic fields.
The international Muon Ionization Cooling Experiment (MICE) at the
Rutherford Appleton Laboratory (UK) has recently demonstrated effective,
$O$(10\%), reduction of the transverse phase-space volume (emittance) of 140\,MeV/c muons passing through a prototype ionization cooling-channel cell
using both lithium-hydride and liquid-hydrogen absorbers within a
magnetic guiding-and-focusing lattice with peak fields of 4\,T ~\cite{Mice2020Nature}.  In the final stages of a cooling channel for a collider, relatively small aperture solenoid magnets having the highest possible fields, tens of Tesla, will be required to deliver beams of the quality required for a multi-TeV collider.  

Technological challenges also arise in other systems required to deliver high quality beams to the collision point in a muon collider. High-gradient normal and superconducting RF systems are required to accelerate the beams rapidly before the beams decay.  Superconducting dipole magnets capable of providing 10 T to 15 T magnetic fields 
are required to keep the collider ring as small as possible, thereby maximizing the number of muon collisions before muon decay dissipates the beams.  In the accelerator ring, fast-ramping magnets will have to be cycled several times a second, which requires that the energy stored in these magnets must be recovered with very high efficiency to preserve the power efficiency of the accelerator complex.  
 
\begin{table}
  \begin{center}
    \begin{tabular}{|l|ccc|}
      \hline
      \hline
      Center of mass energy $\sqrt{s}$ (TeV) & .126 &  3 & 14        \\
      Circumference (km) & 0.3 &  4.5 & 14   \\
      Interaction points & 1 & 2 & 2  \\
      Average luminosity (10$^{34}$ cm$^{-2}$ s$^{-1}$) & 0.008 & 1.8 & 40 \\
      Integrated luminosity/detector (ab$^{-1}$/year) & 0.001 & 0.18 & 4 \\
      Time between collisions~($\mu s$) & 1 & 15 & 47 \\
      Cycle repetition rate~(Hz) & 1 & 5 & 5 \\
      \hline
      Energy spread (rms, \% )& 0.004 & 0.1 & 0.1 \\
      Bunch length (rms, mm)& 63 & 5 & 1 \\
      IP beam size ($\mu$m)&75 & 3.0 & 0.6 \\
      \hline
      Dipole design field (T)  & 10 &  10 & 15 \\
      Proton driver beam power (MW) & 4 & 4 & 1 \\
      Beam power in collider (MW) & 0.08 & 5.3 & 20.2 \\
      \hline
      \hline
    \end{tabular}
  \end{center}
  \caption{
    Tentative parameters of variants of future high-energy muon colliders, including the $\mu^+\mu^-$ Higgs factory.
    A detailed design of a 14 TeV center-of-mass muon collider design is not yet complete and the numbers shown here are a projection \cite{neuffer2018}. 
  }
  \label{MCALL}
\end{table}

Muon decays produce intense neutrino fluxes that are concentrated primarily in line with the straight sections of the collider ring. A small fraction of these neutrinos interact with the rock and other matter as they emerge at the surface of the Earth, thus producing ionizing radiation.  The neutrino interaction rate in the vicinity of the surface rises linearly with energy. The impact of this neutrino-induced radiation can be mitigated, for example, by adding a vertical perturbation in the collider orbit \cite{King}.  A further reduction in the neutrino-induced radiation would be obtained if the emittance of the muon beam was reduced so that the required luminosity could be obtained using a significantly smaller number of muons.  A novel muon production scheme, LEMMA, has recently been proposed in which muon pairs are produced through $e^+e^-$ annihilation just above the production threshold when 45\,GeV positron beam strikes a solid target \cite{LEMMA}. This scheme might allow beams to be produced with much lower current but much higher phase-space density, thus delivering the same luminosity but with significantly reduced neutrino-induced radiation.

Comprehensive overviews of the techniques that have been developed to
address the issues relevant for the construction of a muon collider can be found in references~\cite{palmer2014muonrast,boscolo2018futuremuon}.
These publications, and the documents to which they refer, describe 
the significant progress that has been made and summarise the feasibility studies that have been carried out to demonstrate that a muon collider with a $\sqrt{s} >10$\,TeV can be built with current or emerging technology (see Table \ref{MCALL}). 

The development of an energy-frontier muon collider has elements that have great synergy with other efforts in the field.  For instance, the need for the development of high-field magnets parallels the ongoing R$\&$D program for very high energy proton-proton colliders \cite{FCC}.  The development of a high brightness muon source would also benefit other scientific endeavours.  In particular, muons from a proton driver-based source would provide high purity and precisely characterized neutrino beams for long- and short-baseline neutrino experiments \cite{geer1998,IDS-NF,NuMAX}.


\section{Path To an Energy Frontier Muon Collider}

The LHC luminosity upgrade will extend the physics program at the world's highest energy collider to about 2040.
It is possible to envision a path towards an energy-frontier muon collider in Europe by the mid- to late 2040s. The technically-limited plan starts with an initial four year development phase to establish baseline designs for a 3\,TeV collider with a luminosity of $\sim 2 \cdot 10^{34}$~cm$^{-2}$s$^{-1}$ and a $>$10\,TeV collider with a luminosity of $\sim 4 \cdot 10^{35}$~cm$^{-2}$s$^{-1}$.
The discovery potential of the $>$10\,TeV machine would be competitive to or exceed that of any other energy frontier collider option being evaluated at the moment.

The resulting baseline designs will allow to evaluate the cost scale and risks of a muon collider and define the muon production, cooling, and acceleration test facility (or facilities) as a basis to decide on the future of the project.
The initial phase of the program would be followed by a second phase of roughly 6 years to construct and operate the test facility. The collider design would also be optimized during this period. The results of this second phase would lay the foundations for a decision to move forward into the third four-year phase during which a full technical design would be developed.  The construction of the muon collider itself is estimated to require a further 10 years.

The focus of the technical development towards successful implementation of a muon collider will be on key systems that can reduce the cost of the collider and to increase its power efficiency and performance. Laboratories in Europe, Asia, the US, and around the world have sufficient expertise to deliver elements of the program.  These laboratories are joining together to form the international collaboration required to explore the various options and to develop an integrated design concept that encompasses the physics, the detectors and the accelerator.  
This effort will bring the outstanding features of the muon collider to bear on the exploration of the {\it Terra Promissa} of new phenomena that are beyond reach of the LHC.

\end{document}